\pdfoutput=1
\documentclass[pdftex,twocolumn,epjc3]{svjour3}          

\RequirePackage[T1]{fontenc}

\smartqed  

\RequirePackage{graphicx}
\RequirePackage{mathptmx}      
\RequirePackage{flushend}
\RequirePackage[numbers,sort&compress]{natbib}
\RequirePackage[colorlinks,citecolor=blue,urlcolor=blue,linkcolor=blue]{hyperref}

\journalname{Eur. Phys. J. C}

\usepackage{dcolumn}
\usepackage{color}
\usepackage{booktabs}
\usepackage{xspace}
\usepackage{graphicx}
\usepackage{amsmath}
\usepackage{amssymb}
\usepackage[figuresright]{rotating}
\usepackage{subfigure}
\usepackage{lineno}

\graphicspath{
 {figure/}
}

\def \be {\begin{equation}}
\def \ee {\end{equation}}
\def \ee  {\end{equation}}
\def \bea {\begin{eqnarray}}
\def \eea {\end{eqnarray}}

\begin{document}
\title{
Azimuthal dependence of two-particle transverse momentum current correlations
}
\medskip
\author{Niseem Magdy \thanksref {e1, addr1} \and
Sumit Basu \thanksref {e2, addr2} \and
Victor Gonzalez  \thanksref {addr3} \and
Ana Marin \thanksref {addr4} \and
Olga Evdokimov  \thanksref {addr1} \and
Roy A. Lacey  \thanksref {addr5} \and
Claude Pruneau \thanksref {e3, addr3} 
}
\thankstext{e1}{e-mail: {\em niseemm@gmail.com}}
\thankstext{e2}{e-mail: {\em sumit.basu@cern.ch}}
\thankstext{e3}{e-mail: {\em claude.pruneau@wayne.edu}}

\institute{Department of Physics, University of Illinois at Chicago, Chicago, Illinois 60607, USA\label{addr1} \and 
Lund University, Department of Physics, Division of Particle Physics, Box 118, SE-221 00, Lund, Sweden\label{addr2} \and 
Department of Physics and Astronomy, Wayne State University, Detroit, Michigan 48201, USA\label{addr3} \and
GSI Helmholtzzentrum f\"ur Schwerionenforschung, Research Division and ExtreMe Matter Institute EMMI, Darmstadt, Germany\label{addr4} \and 
Department of Chemistry, State University of New York, Stony Brook, New York 11794, USA\label{addr5}}

\date{Received: date / Revised version: date}
\maketitle

\begin{abstract}
Two-particle transverse momentum correlation f\-unctions are a powerful technique for understanding the dynamics of relativistic heavy-ion collisions. 
Among these, the transverse momentum correlator $G_{2}\left(\Delta\eta,\Delta\varphi\right)$ is of particular interest for its potential  sensitivity to the shear viscosity per unit of entropy density $\eta/s$ of  the quark-gluon plasma formed  in heavy-ion collisions.
We use the UrQMD, AMPT, and EPOS models for Au--Au at $\sqrt{s_{\rm NN}}$ = 200~GeV and Pb--Pb at $\sqrt{s_{\rm NN}}$ = 2760~GeV to investigate the long range azimuthal dependence of  $G_{2}\left(\Delta\eta,\Delta\varphi\right)$, and explore its  utility to constrain $\eta/s$ based on charged particle correlations.
We find that the three models yield quantitatively distinct transverse momentum Fourier harmonics coefficients $a^{p_{\rm T}}_{\rm n}$. We also observe  these coefficients exhibit a significant  dependence on  $\eta/s$ in the context of the AMPT model. These observations suggest that  exhaustive measurements of the dependence of  $G_{2}\left(\Delta\varphi \right)$  with collision energy, system size, collision centrality, in particular, offer the potential to  distinguish between different theoretical models and their underlying assumptions. Exhaustive analyses of $G_{2}\left(\Delta\varphi \right)$ obtained in large and small systems should also be instrumental in establishing new  constraints for precise extraction of $\eta/s$ .
\end{abstract}

\section{Introduction}
A central purpose of the heavy-ion programs at the Large Hadron Collider (LHC) and the Relativistic Heavy-Ion Collider (RHIC) is to determine the properties of  quark-gluon plasma (QGP)~\cite{Shuryak:1978ij,Shuryak:1980tp,Muller:2012zq} created in high-energy heavy-ion collisions (A--A). 
Of specific interest are the transport properties of QGP, particularly the specific shear viscosity, shear viscosity per unit of entropy density,  $\eta/s$, which  characterizes the ability of  QGP to transport and dissipate momentum. Studies of $\eta/s$ have gained broad consideration both theoretically and experimentally~\cite{Shuryak:2003xe,Romatschke:2007mq,Luzum:2008cw,Bozek:2009dw,Acharya:2019vdf,Acharya:2020taj,Adam:2020ymj}. By and large, studies of shear viscosity have so far centrally relied on  hydrodynamical models of the large radial and anisotropic flow experimentally observed in heavy-ion collisions. This flow is driven by  asymmetric  pressure gradients in the overlapping region, known as participants, of the nuclei colliding at finite impact parameter. The pressure gradients drive an asymmetric expansion of the fireball which  eventually translates into anisotropic particle emission in the collision transverse plane. Shear viscosity, however, dampens the development of this anisotropy. 
It is thus commonly considered that models of the system expansion without and with tunable viscous forces may enable a reasonably accurate determination of the magnitude of $\eta/s$ in the QGP~\cite{Danielewicz:1998vz,Ackermann:2000tr,Adcox:2002ms,Heinz:2001xi,Hirano:2005xf,Huovinen:2001cy,Hirano:2002ds,Romatschke:2007mq,Luzum:2011mm,Song:2010mg,Qian:2016fpi,Schenke:2011tv,Teaney:2012ke,Gardim:2012yp,Lacey:2013eia,Magdy:2020gxf}.

Various considerations unfortunately limit the achievable precision from  the comparison of hydrodynamic  model predictions with the flow coefficients measured at RHIC and LHC and estimates of  $\eta/s$ still bear sizable  uncertainties~\cite{Shuryak:2003xe,Romatschke:2007mq,Luzum:2008cw,Bozek:2009dw,Song:2010mg,Shen:2011eg}. Much of the  uncertainties stem, in particular, from the limited knowledge of  the initial-state eccentricity of the participant region. 
 
Several new methods have thus been considered to reduce theoretical and experimental uncertainties and progress towards  more robust extractions of the QGP $\eta/s$ and its dependence with the system temperature $T$, $\eta/s(T)$~\cite{Schenke:2019ruo,Alba:2017hhe,Gonzalez:2020bqm,Everett:2020xug,Gardim:2020mmy,Gonzalez:2020gqg,Agakishiev:2011fs,Acharya:2019oxz,Kovtun:2004de,Luzum:2008cw,Song:2010mg,Bernhard:2019bmu}. 
Although those studies have improved the accuracy of the $\eta/s$ extraction~\citep{Chatrchyan:2013kba,Sirunyan:2019izh,ALICE:2016kpq,Adam:2020ymj,Niemi:2015qia,Danielewicz:2002pu,Luzum:2010fb,Teaney:2010vd,Adams:2005ca,Magdy:2019ojv,Adamczyk:2016gfs,Ollitrault:2009ie,Adam:2019woz,Magdy:2018itt, Adamczyk:2017ird,Magdy:2017kji,Adamczyk:2017hdl,Alver:2010gr,Magdy:2020bij,Adare:2011tg,Adamczyk:2013waa,Acharya:2017zfg},  further constraints are  needed to reduce  $\eta/s$ uncertainties associated with the initial-state ambiguities ~\cite{Song:2010mg,Qiu:2011hf,Song:2012tv} as well as its  dependence on the system's temperature. 

A relatively new strategy for supplementing constraints on $\eta/s$ based on flow measurements  is to leverage the longitudinal and the azimuthal correlations of the transverse momentum two-particle correlator $G_{2}(\Delta\eta,\Delta\varphi)$~\cite{Gavin:2006xd,Sharma:2008qr} defined according to
%
\begin{eqnarray}\label{eq:G2}\nonumber
G_{2}(\Delta\eta,\Delta\varphi) &=&  \int_{\Omega_1,\Omega_2} G_{2}\left(\eta_1,\varphi_1, \eta_2,\varphi_2\right)
\delta(\Delta\eta-\eta_1+\eta_2)\\ &&\times {\rm d}\eta_1 {\rm d}\eta_2
\delta(\Delta\varphi-\varphi_1+\varphi_2){\rm d}\varphi_1{\rm d}\varphi_2,
\end{eqnarray}
where $\Omega_1$, $\Omega_2$, represent the kinematic acceptance of particle 1 and 2, and

\begin{eqnarray} \label{eq:C2}\nonumber
G_{2}\left(\eta_1,\varphi_1, \eta_2,\varphi_2\right)&=& 
\frac{
S_2(\eta_1,\varphi_1,p_{\rm T,i}, \eta_2,\varphi_2,p_{\rm T,j})
}
{
\rho_1(\eta_1,\varphi_1)
\rho_1(\eta_2,\varphi_2)
}\\
&&-
\left\langle p_{\rm T}(\eta_1,\varphi_1)\right\rangle
\left\langle p_{\rm T}(\eta_2,\varphi_2)\right\rangle,
\end{eqnarray}
with
\begin{eqnarray}\nonumber
S_2(\eta_1,\varphi_1, \eta_2,\varphi_2) &=& \int_{\Omega_1,\Omega_2}
\rho_2(\eta_1,\varphi_1,p_{\rm T,1}, \eta_2,\varphi_2,p_{\rm T,2})\\
&& \times p_{\rm T,i} 
p_{\rm T,j}
{\rm d}p_{\rm T,i} 
{\rm d}p_{\rm T,j}, \\ 
\rho_1(\eta_i,\varphi_j) &=& \int_{\Omega_i}
\rho_1(\eta_i,\varphi_i,p_{\rm T,i}) {\rm d}p_{\rm T,i}, \\
\left\langle p_{\rm T}(\eta_i,\varphi_i)\right\rangle &=&
\frac{
\int_{\Omega_i} \rho_1(\eta_i,\varphi_i,p_{\rm T,i}) p_{\rm T,i}{\rm d}p_{\rm T,i}}
{\int_{\Omega_i}
\rho_1(\eta_i,\varphi_i,p_{\rm T,i}) {\rm d}p_{\rm T,i}},
\end{eqnarray}
in which $\rho_1(\eta_i,\varphi_i,p_{\rm T,i})$ and 
$\rho_2(\eta_1,\varphi_1,p_{\rm T,1} , \eta_2,\varphi_2,p_{\rm T,2})$ are single and pair densities computed as 
\begin{eqnarray} \label{eq:density}
\rho_1(\eta_i,\varphi_i,p_{\rm T,i})&=& 
\frac{{\rm d^3}N}{{\rm d}\eta_i{\rm d}\varphi_i {\rm d}p_{{\rm T},i} }, 
\end{eqnarray}
\begin{eqnarray}\nonumber
\rho_2(\eta_1,\varphi_1,p_{\rm T,1}, \eta_2,&&\varphi_2,p_{\rm T,2})=\\
&&\frac{{\rm d^6}N}{{\rm d}\eta_1{\rm d}\varphi_1 {\rm d}p_{{\rm T},1} {\rm d}\eta_2{\rm d}\varphi_2 {\rm d}p_{{\rm T},2}}.
\end{eqnarray} 


The correlator $G_{2}\left(\Delta\eta,\Delta\varphi\right)$ amounts to a measure of the covariance of momentum currents \cite{Gavin:2006xd}. Accordingly, it is sensitive to  dissipative viscous effects unravelling during the transverse and longitudinal expansion of the medium created in heavy-ion collisions. The broadening of its longitudinal width, shown to be sensitive to the magnitude of $\eta/s$~\cite{Gavin:2006xd}, has been observed by both RHIC and LHC experiments~\cite{Sharma:2008qr,Acharya:2019oxz,Magdy:2020fma}. It has even been used to extract a centrality dependence $\eta/s$ value at the two energies~\cite{Gonzalez:2020bqm}. On the other hand, it remains an open question whether  the azimuthal dependence of the transverse momentum correlator $G_{2}\left(\Delta\eta,\Delta\varphi \right)$ also carries information about $\eta/s$. It is thus of interest to examine whether established heavy-collision models such as UrQMD, AMPT, and EPOS can qualitatively, if not quantitatively, reproduce correlation functions reported by the STAR and ALICE collaborations. It is also of interest to examine whether changes in the viscosity $\eta/s$ used in model calculations of $G_{2}$ are readily reflected by changes of the amplitude or shape of this correlator.
Ideally, one should also consider whether $G_2$  provides sensitivity to the   temperature-dependent $\eta/s(T)$ as well as the bulk viscosity $\zeta/s(T)$. However, such studies  are left for future works given they require the use of models  that include transparent and readily tuneable values  of $\eta/s(T)$ and   $\zeta/s(T)$~\cite{Dubla:2018czx, Everett:2020xug}.

In this work, we investigate the azimuthal dependence of the transverse momentum correlator $G_{2}\left(\Delta\eta,\Delta\varphi \right)$ for Au--Au collisions at $\sqrt{s_{\rm NN}}$= 200 GeV and  Pb--Pb $\sqrt{s_{\rm NN}}$=2760 GeV, simulated with the  UrQMD (Ultra relativistic Quantum Molecular Dynamics)~\cite{Bass:1998ca, Bleicher:1999xi, Petersen:2008dd}, AMPT (A Multi-Phase Transport)~\cite{Lin:2004en}, and EPOS~\cite{Drescher:2000ha,Werner:2010aa,Werner:2013tya} models. A similar study was already conducted~\citep{Basu:2020ldt} to establish whether these models can reproduce the number and  transverse momentum correlators  $R_2$ and $P_2$ in Pb--Pb collisions at $\sqrt{s_{\rm NN}}=$ 2760~GeV \cite{Adam:2017ucq,Acharya:2018ddg}. 
Furthermore, we also explore the sensitivity of the azimuthal dependence of the $G_{2}(\Delta\eta, \Delta\varphi)$ correlator to the magnitude of  $\eta/s$  as well as its capacity to  constrain  theoretical models.


This paper is organized as follows. Section~\ref{sec:2} describes details of the analysis method and the theoretical models used to investigate the sensitivity of the $G_2$ correlator to details of the collision dynamics. In Sec.~\ref{sec:3}, calculations of the $G_2$ correlators based on the  UrQMD, AMPT, and EPOS models are reported and discussed.  A summary is presented in Sec.~\ref{sec:4}.

\section{Methodology} \label{sec:2}
We describe the models used in this work in sec.~\ref{sec:2a} and the  analysis techniques used to compute $G_2$  in sec.~\ref{sec:2b}

\subsection{Models}\label{sec:2a}
This study is performed with simulated events of  Au--Au collisions at $\sqrt{s_{\rm NN}}$ = 200~GeV and Pb--Pb at $\sqrt{s_{\rm NN}}$ = 2760 \-GeV, obtained with the
UrQMD~\cite{Bass:1998ca, Bleicher:1999xi, Petersen:2008dd}, AMPT~\cite{Lin:2004en}, and EPOS \-\cite{Drescher:2000ha,Werner:2010aa,Werner:2013tya} models.
The collision dynamics of interest belongs to the medium-bulk  regime.
Computations of $G_2$ are thus limited to particles in the transverse momentum  range $0.2 < p_T < 2.0$ GeV/$c$. Additionally, in order to mimic the acceptance of the STAR experiment at RHIC and the ALICE experiment at LHC, the correlator calculations  are further restricted to $|\eta|$ $<$ $1.0$ and $0.8$, respectively. 

\begin{itemize}
\item{UrQMD Model}:
The UrQMD is a microscopic model that has been widely used to study the ultra-relativistic heavy-ion collisions~\cite{Bass:1998ca,Bleicher:1999xi,Petersen:2008dd}.
It was originally designed to study hadron-hadron, hadron-nucleus, and  heavy-ion collisions from $E_{\rm Lab} = 100$ A$\cdot$MeV to $\sqrt{s_\text{NN}} =$~200 GeV.
It features a collision term accounting for more than 50 baryons (anti-baryons) and 40 mesons (anti-mesons). The UrQMD model describes the hadron-hadron interactions and the system evolution based on covariant propagation of all hadrons in the model with stochastic binary scattering, resonance decay, and color string formation. 
UrQMD was  recently  upgraded and now features a hybrid configuration that describes the evolution of QGP with  an intermediate hydrodynamical stage~\cite{Petersen:2008dd}.
In this work, we used the original parton and hadron transport version (release 3.3) towards the simulations of Au--Au collisions at RHIC whereas the hybrid version (release 3.4) is used for the simulation of Pb--Pb collisions at LHC. Use of these two UrQMD versions, in conjunction with comparisons with results from AMPT and EPOS, enables an assessment of the necessity of hydrodynamics stage at RHIC energies. We will see, indeed, that the original version does not appear to build up the large amount of flow observed in Au--Au collisions at RHIC while the hybrid version somewhat overshoots the $v_2$ and $v_3$ coefficients reported by the ALICE collaboration.


\item{AMPT Model}: The AMPT  model (v2.26t9b)~\cite{Lin:2004en} has been extensively used to study relativistic heavy-ion collisions  at RHIC and LHC energies. It is found to successfully reproduce several of the observables measured in A--A collisions in both these energy ranges ~\cite{Lin:2004en,Ma:2016fve,Solanki:2012ne,Basu:2016dmo,Bhaduri:2010wi,Xu:2010du,Magdy:2020bhd,Guo:2019joy}. 
AMPT nominally provides several optional mechanisms. 
 In this work, we compute Au--Au and Pb--Pb collisions with the  string melting option known to favor the build up of both radial and anisotropic flow.
Key components of AMPT include (i) an initial parton-production stage based on  the HIJING model~\cite{Wang:1991hta,Gyulassy:1994ew}, (ii) a parton scattering stage, (iii) hadronization through coalescence  followed (iv)  by a hadronic interaction stage~\cite{Li:1995pra}.  
The parton scattering  cross-sections used in stage (ii) are estimated according to 
\begin{eqnarray} \label{eq:21}
\sigma_{pp} &=& \dfrac{9 \pi \alpha^{2}_{s}}{2 \mu^{2}},
\end{eqnarray}
where $\alpha_{s}$ is the QCD coupling constant and $\mu$ is the screening mass in the partonic matter. They largely define the expansion dynamics of A--A collision systems \cite{Zhang:1997ej}; 
Within the context of AMPT, the nominal $\eta/s$ magnitude   can be modified via an appropriate selection of $\mu$ and/or $\alpha_s$ for a particular initial temperature $T_{i}$~\cite{Xu:2011fi,Solanki:2012ne}.
\begin{eqnarray} \label{eq:22}
 \dfrac{\eta}{s} &=& \dfrac{3 \pi}{40 \alpha^{2}_{s}}  \dfrac{1}{ \left(  9 +  \dfrac{\mu^2}{T^2} \right)  \ln\left(\dfrac{18 + \mu^2/T^2}{ \mu^2/T^2 } \right) - 18},
\end{eqnarray}
In this work, our simulations of Au--Au collisions at 
$\sqrt{s_{\rm NN}}$ = 200 GeV are performed with ampt-v2.26t9b  at a fixed value  $\alpha_{s}$ = 0.47 but the shear viscosity $\eta/s$ is varied over the range 0.1--0.3 by tuning  $\mu$ from  2.26 to 4.2~$fm^{-1}$ for a temperature  $T_{i}$ = 378 MeV~\cite{Xu:2011fi}. 
Additionally, the simulation of Pb--Pb collisions at $\sqrt{s_{\rm NN}}=$ 2760~GeV are performed with version ampt-v1.26t7-v2.26t7 at a fixed values of  $\alpha_{s}$ = 2.265 and  $\mu$ = 0.33~$fm^{-1}$~\cite{Basu:2020ldt}.

\item{EPOS Model}: 
The event generator EPOS~\cite{Drescher:2000ha,Werner:2010aa,Werner:2013tya} is  bas\-ed on a 3+1D viscous hydrodynamical representation of A--A collisions. The initial state conditions are described in terms of  flux tubes  computed based on Gribov-Regge multiple scattering theory~\cite{Drescher:2000ha}. Three EPOS features are of particular interest in the study of correlation functions:
(i) Division of initial state flux tubes  into \textit{core} and \textit{corona} components  based on the probability that a particle can escape from the ``bulk matter". This division   depends on the fragment transverse momentum and the local string density. The progressive evolution of the latter insures a realistic growth of the strangeness production with increasing centrality as well as a seamless evolution of correlation functions with collision centrality.

(ii)  An hydrodynamical evolution based on the 3D+1 hydrodynamics (i.e. viscous HLLE-based algorithm (vHLLE)) which is itself based on a realistic Equation of State  compatible with Lattice QCD data \cite{Allton:2002zi}.
(iii) A hadronic cascade \textit{hadronic afterburner}  based on components of the UrQMD model~\cite{Bleicher:1999xi,Bass:1998ca} meant to provide a realistic simulation of the role of the short lived post-QGP hadron phase.

\end{itemize}
The correlation functions reported in sec.~\ref{sec:2b}, were obtained for minimum bias events Au--Au collisions at $\sqrt{s_{\rm NN}}=$ 200 GeV and Pb--Pb collisions at $\sqrt{s_{\rm NN}}=$ 2760 GeV.  UrQMD and AMPT data sets were produced by these authors whereas the EPOS event sets were generated and provided  by K. Werner et al.~\cite{Stefaniak:2020oxd,Werner:2013tya}. 
A total of 2.0, 5.0, and 0.35M Au--Au and 0.34, 0.2, and 0.32~M Pb--Pb minimum bias events were generated with UrQMD, AMPT, and EPOS, respectively. 

\subsection{Analysis Method} \label{sec:2b}

The minimum bias event data sets produced with the UrQMD, AMPT, and EPOS models were partitioned into several classes of collision centrality based on the  impact parameter of the collisions. Simulated events were used to study the $G_2$ correlator, based on  Eq.~(\ref{eq:C2}), as well as the collision centrality dependence of the  strength of the elliptic and triangular flow harmonics $v_{2}$ and $v_{3}$, respectively.  Below, we describe  the methods used to compute  the $G_2$ correlator and determine the $v_{2}$ and $v_{3}$  harmonic coefficients.

\subsubsection{The $G_2$ correlator}

The correlator $G_2$, defined in  Eq.~(\ref{eq:C2}), was computed in each centrality class, based on the number of particles observed  event-by-event, according to 
\begin{eqnarray} \label{eq:23}
G_{2}\left(\eta_1,\varphi_1,\eta_2,\varphi_2 \right)  &=& 
\frac{
\left\langle \sum\limits_{\text{i}}^{n_1} \sum\limits_{\text{j} \neq \text{i}}^{n_{2}}  p_{\rm T,i}p_{\rm T,j} \right\rangle
}
{\langle n_{1} \rangle \langle n_{2} \rangle} \\ \nonumber
&&-
\left\langle p_{\rm T,1}\right\rangle_{\eta_1,\varphi_1}
\left\langle p_{\rm T,2}\right\rangle_{\eta_2,\varphi_2}
\end{eqnarray}
where $n_1\equiv n(\eta_1,\varphi_1)$ and $n_2\equiv n(\eta_2,\varphi_2)$ are  event-wise multiplicities of charged 
particles in bins $\eta_1,\varphi_1$ and $\eta_2,\varphi_2$ respectively; $p_{T,i}$ and $p_{T,j}$ are the transverse momenta of particles i$^{th}$ and j$^{th}$ in their respective bins; and $\langle O\rangle$ represents  an event-ensemble  average  of the quantity $O$.
More extensive descriptions  of the $G_{2}$ correlation function  and its properties are presented in Refs.~\citep{Gavin:2006xd,Sharma:2008qr,Magdy:2020fma}.

The $G_{2}$($\Delta\eta$,$\Delta\phi$) correlators studied in this work were  first constructed as functions of  $\Delta\eta$ and $\Delta\phi$ using  40- and 60-bins, respectively. However, given  our specific interest on the azimuthal dependence of $G_2$ for large pseudorapidity gaps (i.e. long range behavior), we used a pseudorapidity gap requirement of $|\Delta\eta| > 0.7$ and projected $G_2$ correlation functions onto the $\Delta\phi$ axis. The selection of this specific $\eta$-gap was in part motivated by observations by the ALICE collaboration~\cite{Acharya:2018ddg} which reported that short-range correlations become essentially negligible beyond $|\Delta\eta| \gtrapprox 0.7$.

Fourier decompositions of the  $G_{2}$($\Delta\phi$) correlator projections were computed for each collision centrality class  using the fit function 
\begin{eqnarray} \label{eq:28}
f\left(\Delta\varphi \right) &=& a^{p_{\rm T}}_{\rm 0} + 2~\sum\limits_{\text{n=1}}^{\text{6}} A^{p_{\rm T}}_{\rm n} cos(n~\Delta\varphi),
\end{eqnarray}
and the flow-like coefficients $a^{p_{\rm T}}_{\rm n}$ were computed according to 
\begin{eqnarray} \label{eq:28b}
a^{p_{\rm T}}_{\rm n} &=& A^{p_{\rm T}}_{\rm n}/\sqrt{|A^{p_{\rm T}}_{\rm n}|}.
\end{eqnarray}
Nominally, the coefficients $A^{p_{\rm T}}_{\rm n}$ may be either negative,  positive, or null. We found, however, that fit values obtained from $G_2$ correlators computed, in this work, with the UrQMD, AMPT, and EPOS models were always non-negative. 

 \subsubsection{Flow coefficients $v_{n}$}
 
 The flow coefficients, $v_{n}$, were  computed based on the two-particle cumulant technique using the sub-event method presented in Refs.~\cite{Bilandzic:2010jr,Bilandzic:2013kga,Jia:2017hbm,Gajdosova:2017fsc}.  The sub-event method is used with an $\eta$-gap $>$ 0.7 to reduce  non-flow correlations arising from resonance decays, Bose-Einstein correlations, as well as contributions from jet constituents. Particles from each event were grouped into two sub-events $A$ and $B$ belonging to two non-overlapping $\eta$-interval with $\eta_{A}~ > 0.35$ and $\eta_{B}~ < -0.35$, and the flow coefficients were computed according to 
\begin{eqnarray}\label{eq:2-4}
v_{n} &=&  \langle  \langle \cos (n (\varphi^{A}_{1} -  \varphi^{B}_{2} )) \rangle  \rangle^{1/2}.
\end{eqnarray}

Flow harmonic coefficients $v_2$ and $v_3$, discussed in  sec.~\ref{sec:3}, were obtained from the events  produced with UrQMD, AMPT, and EPOS, for particles  within the kinematic range  $|\Delta\eta| > 0.7$, and $0.2<p_{T}<2.0$ GeV/$c$ to match measurements  of these coefficients   by the STAR~\cite{Adam:2019woz} and ALICE~\cite{Acharya:2018lmh} experiments.
The STAR measurements~\cite{Adamczyk:2016exq,Adamczyk:2017hdl} were conducted for Au--Au collisions at $\sqrt{s_\text{NN}} =$~200~GeV with $|\eta| < 1.0$, $|\Delta\eta| > 0.7$, and $0.2<p_{T}<2.0$ GeV/$c$, whereas the ALICE measurements~\cite{Acharya:2018ddg} were obtained based on Pb--Pb collisions at $\sqrt{s_\text{NN}} =$~2760~GeV with $|\eta| < 0.8$, $|\Delta\eta| > 0.9$, and $0.2<p_{T}<2.0$~GeV/$c$.

\section{Results and discussion}\label{sec:3}

\begin{figure}[hbt] 
    \includegraphics[width=1.0 \linewidth, angle=-0,keepaspectratio=true,clip=true]{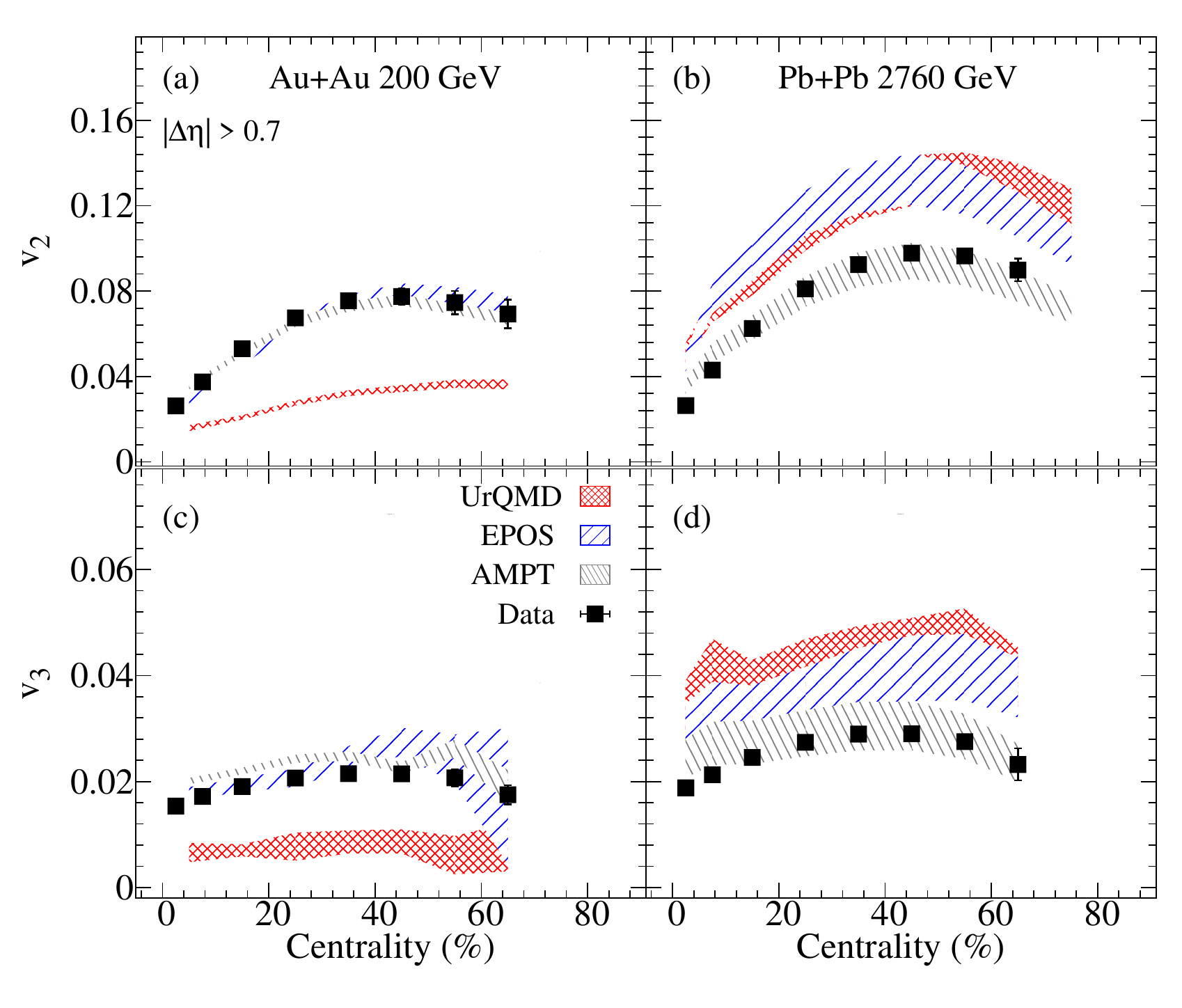}
    \vskip -0.4cm
    \caption{Centrality dependence of the  harmonic coefficients $v_{n}$, $n = 2, 3$,  computed with   UrQMD, AMPT and EPOS   for Au--Au collisions at $\sqrt{s_\text{NN}} =$~200~GeV in panels (a,c) and for Pb--Pb collisions at $\sqrt{s_\text{NN}} =$~2760~GeV in panels (b,d). The solid points are the experimental data reported by STAR~\cite{Adamczyk:2016exq,Adamczyk:2017hdl} and ALICE~\cite{Acharya:2018ddg} whereas the shaded areas represent the $v_{n}$ values obtained in this work.
  } \label{fig:1}
\end{figure}
\begin{figure*}[hbt]
\centering
    \includegraphics[width=0.8 \linewidth, angle=-0,keepaspectratio=true,clip=true]{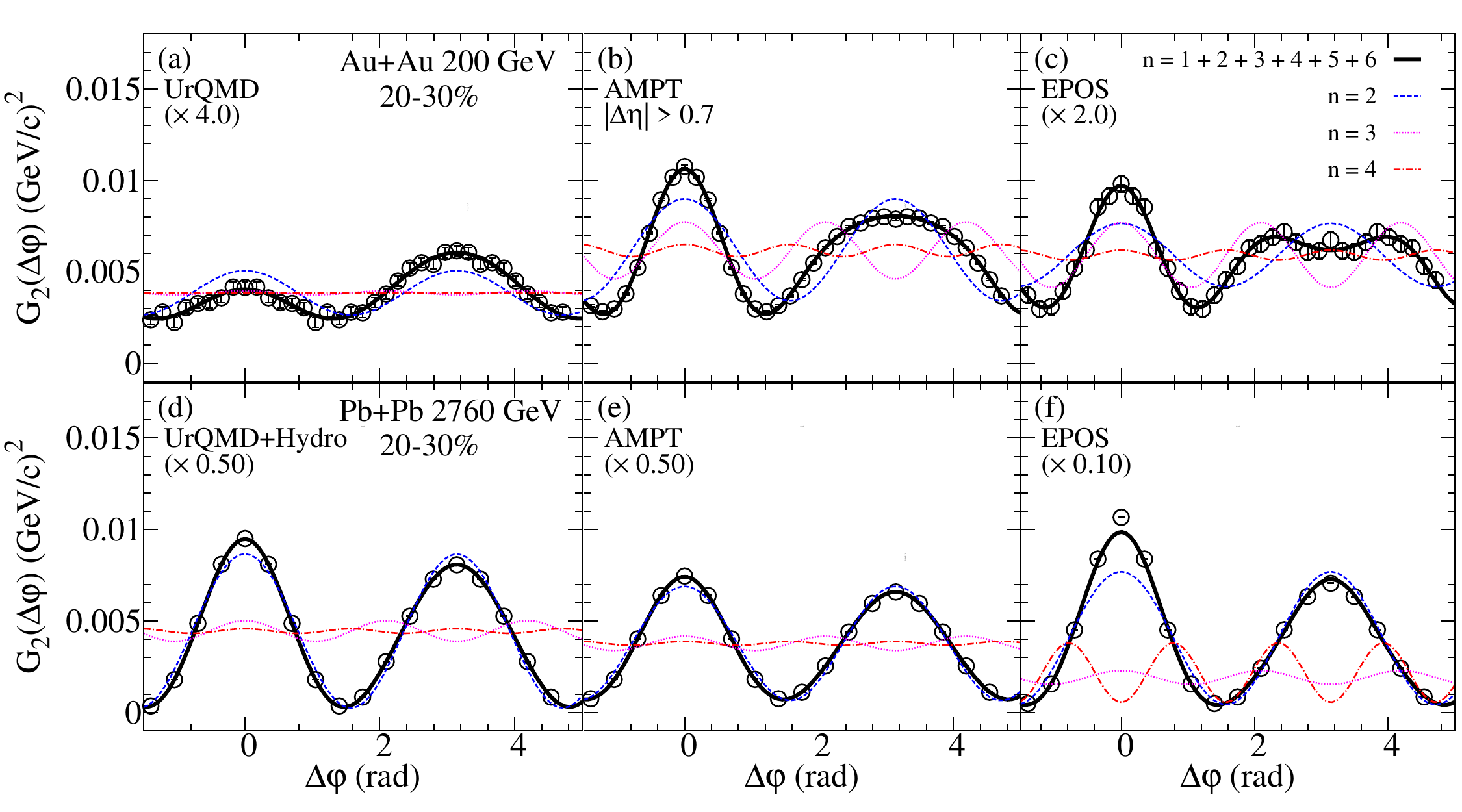}
      \vskip -0.4cm
    \caption{    
  Comparison of the azimuthal two-particle transverse momentum correlation function $G_{2}\left(\Delta\varphi \right)$ with a pseudorapidity gap, $\Delta\eta > 0.7$, obtained from 20-30\%  central UrQMD, AMPT and EPOS events for Au--Au collisions at $\sqrt{s_\text{NN}} =$~200~GeV in panels (a--c) and for Pb--Pb collisions at $\sqrt{s_\text{NN}} =$~2760~GeV in panels (d--f). Solid curves show  Fourier fits to the simulated data with Eq.~\ref{eq:28} and dashed lines show the $n=2,3,4$ components of these fits. In panels  (a,c-f), the correlator amplitudes were scaled by the factors shown for convenience of presentation and comparison of the results obtained with the three models.}  \label{fig:2}
\end{figure*}
\begin{figure}[ht] 
    \includegraphics[width=1.0 \linewidth, angle=-0,keepaspectratio=true,clip=true]{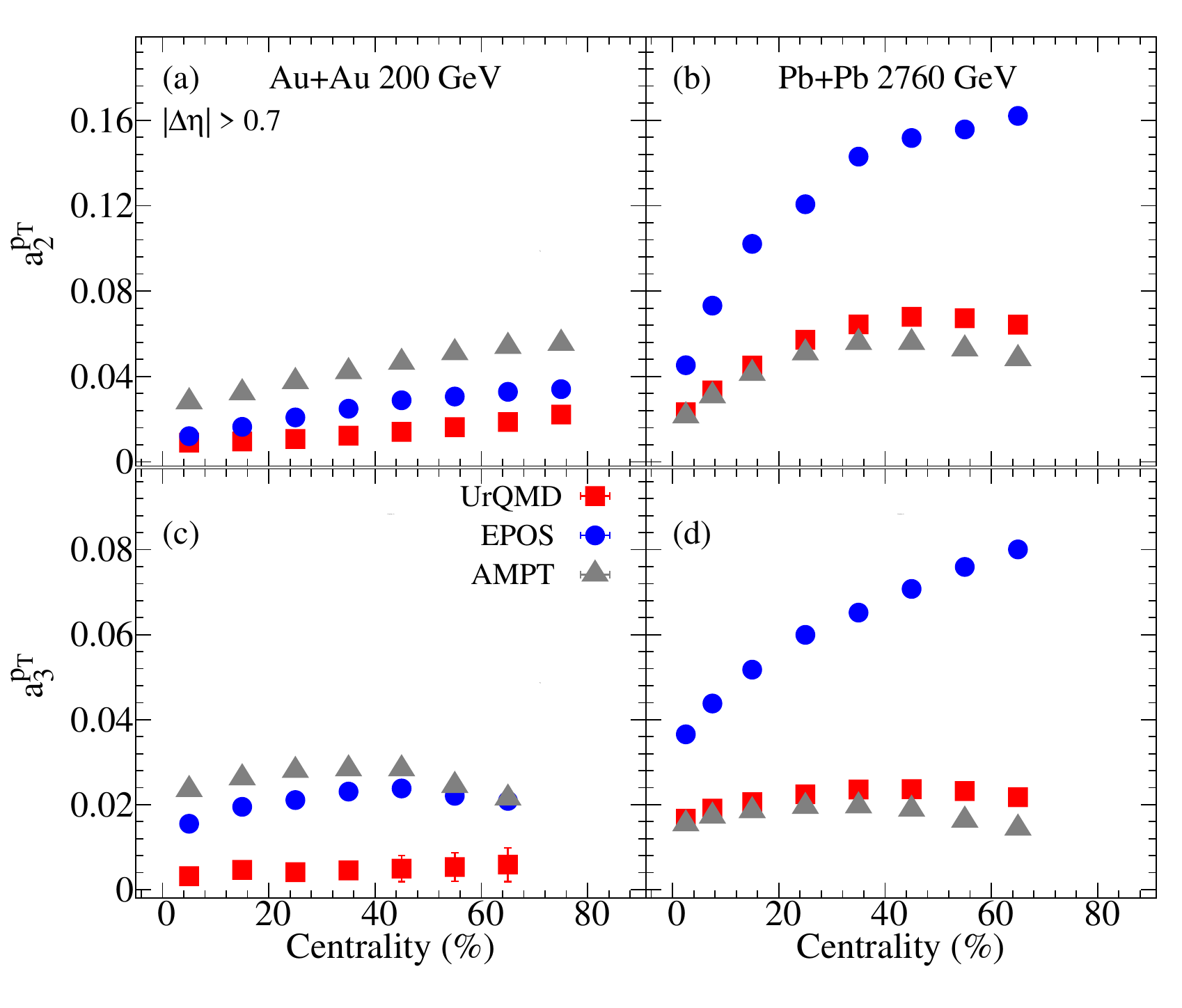}
    \vskip 0.5cm
    \caption{
    Centrality dependence of the coefficients $a^{p_{T}}_{n}$, $n = 2,3$, extracted with  UrQMD, AMPT and EPOS events for Au--Au collisions at $\sqrt{s_\text{NN}} =$~200~GeV in panels (a,c) and for Pb--Pb collisions at $\sqrt{s_\text{NN}} =$~2760~GeV in panels (b,d). 
  } \label{fig:3}
    \vskip -0.2cm
\end{figure}
We compare the collision centrality dependence of the $v_{2}$ and $v_{3}$ coefficients obtained with  the three models with measurements reported by  STAR  and ALICE collaborations~\cite{Adam:2019woz,Acharya:2018lmh} in Fig.~\ref{fig:1}.
We find that the AMPT and EPOS models  quantitatively reproduce both the magnitude and collision centrality evolution of the $v_{2}$ and $v_{3}$ coefficients reported by STAR for Au--Au collisions:  the coefficients are somewhat large in quasi-peripheral collisions (70\% centrality bin), rise to maximum values in the centrality range 40-50\%, and decrease monotonically towards zero in most  central collisions. We note, however, that UrQMD tends to grossly underestimate the magnitude of both  the $v_{2}$ and $v_{3}$ coefficients reported by STAR. 

The UrQMD version (version 3.3) used in this work to simulate Au--Au collisions features only hadron collisions and transport which, as presented, can not reproduce the strength of the $v_n$ observed in Au--Au at RHIC. We thus conclude, in agreement with results reported in prior studies~\cite{Zhu:2005qa,Adamczyk:2012ku}, that the hadron transport implemented in  UrQMD 3.3 is insufficient to account for the magnitude of the $v_n$ coefficients observed experimentally. 

Turning our attention to the Pb--Pb collision  data sets, we find that all three models qualitatively reproduce the magnitude and collision centrality evolution of the $v_2$ and $v_3$ coefficients reported by the ALICE collaboration. We note, however, that AMPT has best success in reproducing the coefficients magnitude while both UrQMD and EPOS overestimate the $v_n$ by approximately 25\% and 30\%, respectively, over the entire collision centrality range reported by ALICE.
The better performance of UrQMD at $\sqrt{s_\text{NN}} =$~2760~GeV seems at odds with its performance in Au--Au collisions at $\sqrt{s_\text{NN}} =$~200 GeV. Note, however, that the hybrid UrQMD version used in  our simulation of Pb--Pb collisions  at $\sqrt{s_{\rm NN}}$\- = 2760~GeV involves a QGP stage described with hydrodynamic evolution. We thus find, again in agreement with prior studies~\cite{Basu:2020ldt}, that the addition of this QGP hydrodynamic stage provides for an increased anisotropic flow build up while the nominal version of UrQMD, which involves only hadron collisions, does not.

The  data-model comparisons shown in Fig.~\ref{fig:1} and prior studies~\cite{Zhu:2005qa,Adamczyk:2012ku,Magdy:2020fma,Basu:2020ldt}, indicate that different theoretical models, with different initial conditions and different values of $\eta/s$,  can describe, to a very good degree of accuracy,   anisotropic flow measurements reported by RHIC and LHC experiments. Comparisons of the measurements of the collision centrality evolution of the $v_2$ and $v_3$ coefficients with model predictions do not provide  sufficient discriminant power to favor either of the models. It is consequently of interest to explore whether other observables, and specifically the $G_2$ correlator, can provide such discriminant. 

We thus turn our attention to the the azimuthal dependence of the $G_{2}\left(\Delta\varphi \right)$ correlator, computed with a large pseudorapidity gap,  $|\Delta\eta| > 0.7$, obtained for 20--30\% central collisions from the UrQMD, AMPT, and EPOS models, sho\-wn in Fig.~\ref{fig:2}. Results are presented for Au--Au at $\sqrt{s_\text{NN}}$ = 200 GeV in   panels (a--c)  and for Pb--Pb at $\sqrt{s_\text{NN}} =$~2760~GeV in panels (d--f). 

The $G_{2}\left(\Delta\varphi \right)$ correlation functions  computed with UrQMD, AMPT, and EPOS  exhibit qualitatively similar dependences on $\Delta\varphi$. The $G_{2}\left(\Delta\varphi \right)$ correlators obtained in 200 GeV Au -- Au and 2760 GeV Pb -- Pb collisions with AMPT and EPOS, as well as the $G_{2}$ computed at 2760 GeV with UrQMD exhibit  strong $\cos (2\Delta\varphi)$ modulations and evidence of higher harmonics commonly associated with collective flow aniso\-tropy. We determine the Fourier components based on fits of $G_2\left(\Delta\varphi \right)$ with Eq.~(\ref{eq:28}) in all centrality classes and plot their evolution  with  centrality  in Fig.~\ref{fig:3}.
We observe that although the coefficients $a^{p_{\rm T}}_{\rm n}$ extracted from the three models show a qualitatively  similar centrality dependence, they in fact exhibit substantial quantitative differences. 

Figure~\ref{fig:3} indicates that the coefficients $a^{p_{\rm T}}_{\rm n}$ are described rather differently by the three  models used in this work. This observation implies that $a^{p_{\rm T}}_{\rm n}$ are sensitive to the underlying physics assumptions and transport mechanisms implemented in these models. 
Consequently, one concludes that  detailed $G_{2}\left(\Delta\varphi \right)$ measurements should provide  useful discriminatory power to test the performance of these and other theoretical models.

Based on the construction of the $G_2$ correlator, one expects its azimuthal Fourier harmonics $a^{p_{\rm T}}_{\rm n}$ should be correlated to the initial spatial anisotropy of the colliding systems.
The degree of such correlation
can be tested using the  Event Shape Engineering (ESE) technique~\cite{Adler:2002pu}. ESE reflects the observation that event-by-event fluctuations of the anisotropic flow coefficient $v_{n}$ (for a fixed centrality), is sizable~\cite{Abelev:2012di}.  Thus,  selections on the magnitude of such fluctuations can be leveraged to influence the magnitude of the $v_n$ and $a^{p_{\rm T}}_{\rm n}$ for a fixed centrality selection.

It is noteworthy that there are two caveats to the ESE technique. First, the selective power of the $q_{2}$ (see Eq. \ref{eq:q2}) selection depends on the magnitude of $v_{2}$ and the event multiplicity. Therefore,  the utility of the method is handicapped by weak flow magnitudes and small event multiplicities~\citep{Bzdak:2019pkr}. Second, non-flow effects, such as resonance decays, jets, etc.~\citep{Voloshin:2008dg}, could potentially bias the $q_{2}$ measurements. However, as suggested earlier, such a bias can be minimized via a $\Delta\eta$ separation between the sub-events used for the evaluation of $q_{2}$ and $v_n$.
%
\begin{figure}[hbt]
  \vskip 0.2cm
\includegraphics[width=1.0 \linewidth,angle=-0,keepaspectratio=true,clip=true]{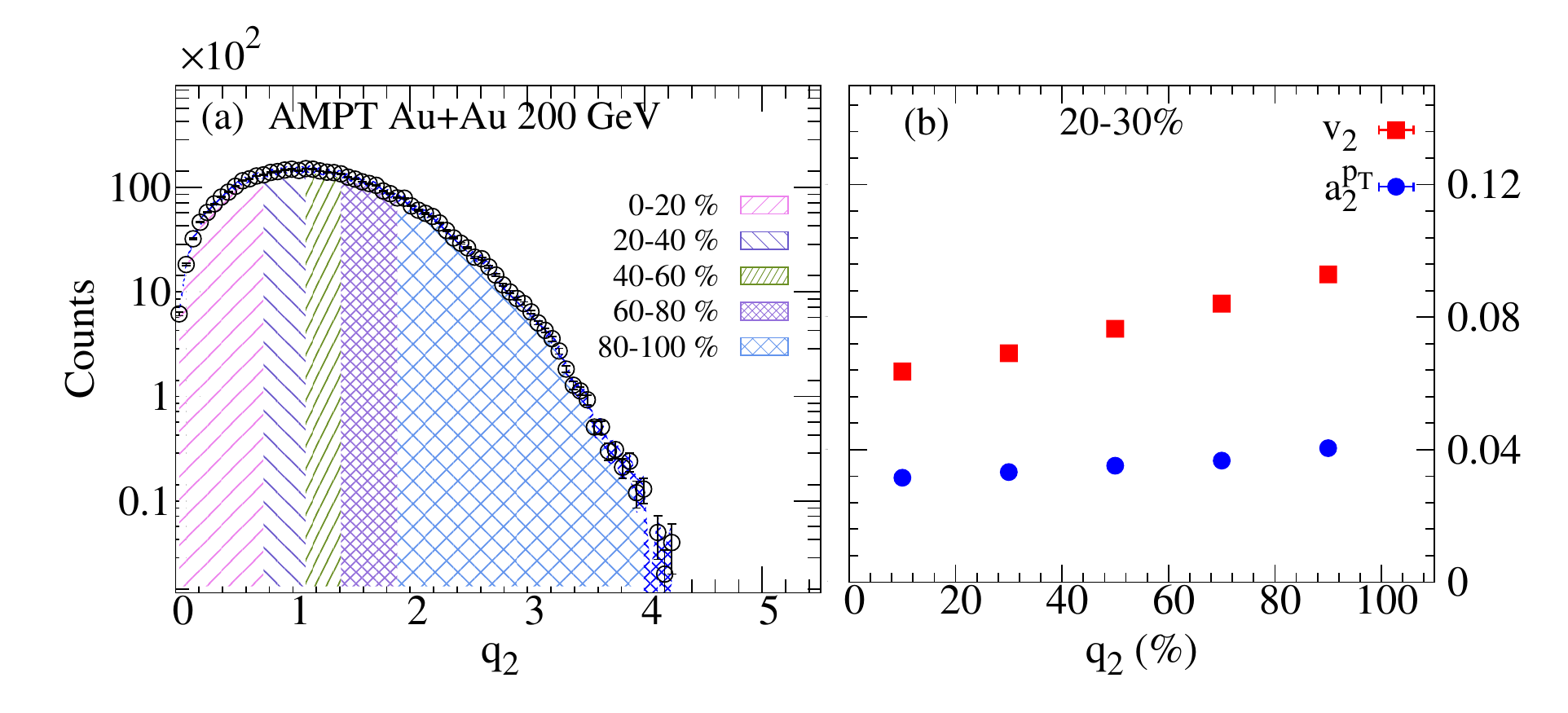}
  \vskip -0.5cm
 \caption{
 (a) Distribution of $q_{2}$  in simulations of $20-30\%$ Au--Au collisions at $\sqrt{s_\text{NN}} =$~200~GeV  with the AMPT model. Shaded areas shown in the left panel identify fractional cross section ranges of  $q_2$ used towards the computations of the evolution of the  $v_2$ and $a_2^{p_{\rm T}}$ coefficients with $q_2$ shown in  panel (b). 
 } \label{fig:4} 
\end{figure}
The event-shape selections were performed via a fractional cut on the distribution of the magnitude of the reduced second-order flow vector, $q_2$~\cite{Schukraft:2012ah,Adler:2002pu}. The flow vector normalized magnitude $q_2$ is computed according to \begin{eqnarray}\label{eq:q2}
q_{2}    &=& \frac{|{Q}_{2}|}{\sqrt{M}}, ~|Q_{2}|  = \sqrt{Q_{2, x}^2 + Q_{2, y}^2}
\end{eqnarray}
with 
\begin{eqnarray}
Q_{2, x} = \sum_{i} \cos(2 \varphi_{i}),
\quad Q_{2, y} = \sum_{i} \sin(2 \varphi_{i}),
\end{eqnarray}
where $|Q_{2}|$ is the magnitude of the second-order harmonic flow vector calculated from the azimuthal distribution of particles within $|\eta| < 0.3$, and $M$ is the charged hadron multiplicity of the same sub-event. Note that the associated flow measurements are performed within $|\eta| > 0.35$ which allows for a separation between the $q_{2}$ subevent and the flow measurements subevents.

Figure~\ref{fig:4} (a) shows  the $q_{2}$ distributions obtained with 20--30\%  Au--Au collision centralities and the $q_2$ based  sub-sample selection  of events used to compute the magnitude of $ v_{2}$ and $a^{p_{T}}_{2}$ coefficients shown in Fig.~\ref{fig:4} (b).
Both $v_{2}$ and $a^{p_{T}}_{2}$ feature  an approximately  linear dependence on  the magnitude of  $q_{2}$ thereby indicating their sensitivity  to the initial eccentricity and eccentricity fluctuations. One notes, however, that the slope ${\rm d}a^{p_{T}}_{2}/{\rm d}q_2$ is considerably smaller than the slope ${\rm d}v_{2}/{\rm d}q_2$ owing most likely to the  different intrinsic dependencies of $a^{p_{T}}_{2}$ and $v_{2}$ on $\left\langle  p_{\rm T}  \right\rangle$~\citep{Adam:2017ucq}. As such, this difference provides a useful additional powerful constraint in the tuning of models and estimations of viscous effects~\citep{Adam:2017ucq,Sharma:2008qr}.

\begin{figure}[hbt]
  \vskip 0.2cm
    \centering
\includegraphics[width=1.0 \linewidth, angle=-0,keepaspectratio=true,clip=true]{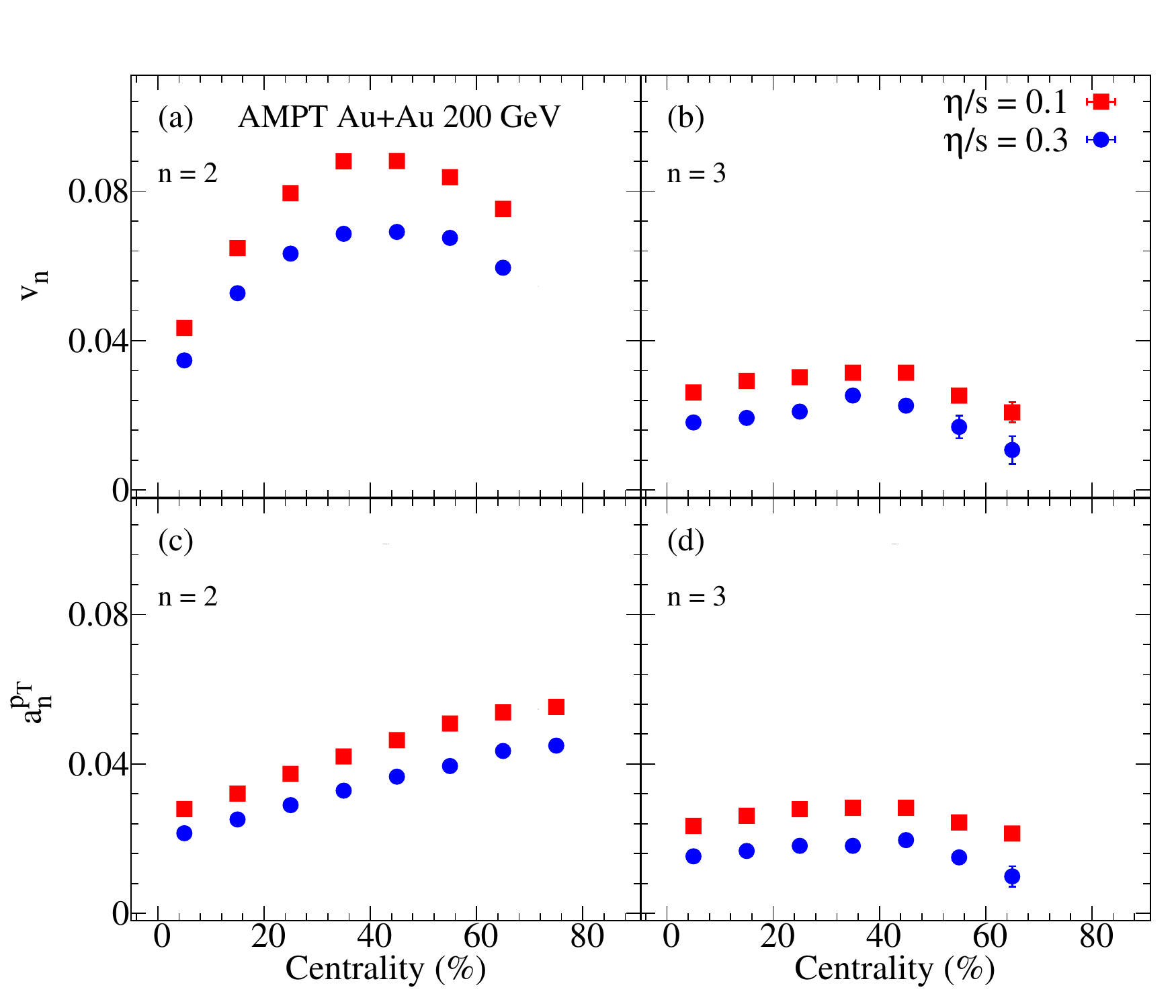} 
\vskip -0.4cm
     \caption{
         Centrality dependence of the coefficients $v_{\rm n}$ (top) and  $a^{p_{\rm T}}_{\rm n}$ (bottom), for  $n = 2$ (left) and $n=3$ (right),
obtained with Au--Au events at $\sqrt{s_\text{NN}} =$~200~GeV generated with  AMPT for  two distinct values of  $\eta/s$. }
    \label{fig:5} 
\end{figure}

The AMPT model was employed in our study of the influence of $\eta/s$ on the azimuthal two-particle transverse momentum correlation function $G_{2}\left(\Delta\varphi \right)$. For these simulations, $\mu$ was varied [with $\alpha_{s}$ = 0.47 and $T_{i}$ = 378 MeV] in conjunction with Eq.~\ref{eq:22} to obtain simulated results for $\eta/s$ =0.1, and 0.3.
Figure~\ref{fig:5} illustrates the centrality dependence of the $v_{2}$ and $a^{p_{T}}_{2}$ coefficients obtained with  $\eta/s$ =0.1, and 0.3 in simulations of  Au--Au collisions at 200 GeV. We find  that
$v_{2}$ and $a^{p_{T}}_{2}$ show a clear sensitivity to the magnitude of $\eta/s$, as well as the expected decrease in the magnitude of $ v_{2}$ and $a^{p_{T}}_{2}$ when $\eta/s$ is increased. 
The observed sensitivity of  $a^{p_{T}}_{2}$ to the magnitude of $\eta/s$  suggests that experimental studies of the $G_{2}\left(\Delta\varphi \right)$  correlator  should provide additional constraints towards precision extraction of $\eta/s$.
However, whether measurements of $G_2$ would also exhibit sensitivity to the temperature-dependence of   $\eta/s$ or the specific bulk viscosity, $\zeta/s$, remains an open question beyond the scope of this study that shall  be investigated in future works~\cite{Dubla:2018czx,Everett:2020xug,Schenke:2020mbo}.

\section{Conclusion} \label{sec:4}
 We presented studies of  the azimuthal dependence of two-particle transverse momentum correlation function $G_{2}\left(\Delta\varphi \right)$ based on Au--Au and Pb--Pb collision simulations with  the UrQMD, AMPT and EPOS models.
We find that the collision centrality dependence of $v_{\rm n}$ flow coeficients obtained with the UrQMD, AMPT, and EPOS models are in qualitative agreement with those observed experimentally by the STAR and ALICE collaborations. 
We note, however, that $a^{p_{\rm T}}_{\rm n}$ centrality dependence is qualitatively similar between these models while the $a^{p_{\rm T}}_{\rm n}$ magnitudes are different, showing the EPOS model an additional agreement between $a^{p_{\rm T}}_{\rm 2}$ and $a^{p_{\rm T}}_{\rm 3}$ up to 30\% central collisions. 
We additionally tested the degree of correlation between $a^{p_{\rm T}}_{\rm n}$ and eccentricity (eccentricity fluctuations) using  the ESE technique which indicated that $a^{p_{T}}_{2}$ increase linearly with the $q_{2}$, and its magnitude is smaller than  $ v_{2}$. The AMPT model with several $\eta/s$ values was used to confirm the $a^{p_{T}}_{2}$ sensitivity to the $\eta/s$ variations.
Based on our UrQMD, AMPT, and EPOS models calculations, we conclude that precise measurements of the azimuthal dependence of $G_{2}\left(\Delta\varphi \right)$ correlator and its collision centrality, system-size and beam-energy dependence will offer new useful tools to test and challenge the theoretical models and can serve as an additional constraint for precision $\eta/s$ extraction.

\section*{Acknowledgments}
The authors  thank Marysia Stefaniak, Jinjin Pan, and Anders Knospe  for  useful discussions.
SB acknowledge the support of the Swedish Research Council (VR). 
This research is supported by the US Department of Energy, Office of Nuclear Physics (DOE NP),  under contracts DE-FG02-94ER40865 (NM and OE), DE-FG02-87ER40331.A008 (RL) and DE-FG02-92ER40713 (CP and VG).


\bibliographystyle{spphys}
\interlinepenalty=10000
\bibliography{ref} 
\end{document}